# Direct Linearly-Polarised Electroluminescence from Perovskite Nanoplatelet Superlattices

Junzhi Ye [1,2], Aobo Ren [3,4], Linjie Dai [1,*], Tomi Baikie [1], Renjun Guo [5], Debapriya Pal [6], Sebastian Gorgon [1], Julian E. Heger [5], Junyang Huang [1], Yuqi Sun [1], Rakesh Arul [1], Gianluca Grimaldi [6], Kaiwen Zhang [7], Javad Shamsi [1], Yi-Teng Huang [1], Hao Wang [8], Jiang Wu [4], A. Femius Koenderink [6], Laura Torrente Murciano [7], Matthias Schwartzkopf [9], Stephen V. Roth [9,10], Peter Müller-Buschbaum [5,11], Jeremy J. Baumberg [1], Samuel D. Stranks [1,7], Neil C. Greenham [1], Lakshminarayana Polavarapu [12], Wei Zhang [3], Akshay Rao [1,*], Robert L. Z. Hoye [2,13,*]

1. Cavendish Laboratory, University of Cambridge, JJ Thomson Ave, Cambridge CB3 0HE, United Kingdom
2. Inorganic Chemistry Laboratory, University of Oxford, South Parks Road, Oxford OX1 3QR, United Kingdom
3. Advanced Technology Institute, University of Surrey, Guildford, United Kingdom
4. Institute of Fundamental and Frontier Sciences, University of Electronic Science and Technology of China, Chengdu, China
5. Lehrstuhl für Funktionelle Materialien, Physik-Department, Technische Universität München, James-Franck-Str. 1, 85748 Garching, Germany
6. Centre for Nanophotonics, AMOLF, 1098 XG Amsterdam, The Netherlands
7. Department of Chemical Engineering and Biotechnology, University of Cambridge, Cambridge CB3 0AS, United Kingdom.
8. Division of Electrical Engineering, Department of Engineering, University of Cambridge, Cambridge, United Kingdom
9. Deutsches Elektronen-Synchrotron (DESY), Notkestrasse 85, D-22607 Hamburg, Germany
10. Department of Fibre and Polymer Technology, KTH Royal Institute of Technology, Teknikringen 56–58, Stockholm, 10044 Sweden
11. Heinz Maier-Leibnitz Zentrum (MLZ), Technische Universität München, Lichtenbergstr. 1, D-85748 Garching, Germany
12. CINBIO, Universidade de Vigo, Materials Chemistry and Physics Group, Department of Physical Chemistry, Campus Universitario As Lagoas, Marcosende, 36310 Vigo, Spain
13. Department of Materials, Imperial College London, Exhibition Road, London SW7 2AZ, United Kingdom

* Email: ld474@cam.ac.uk (L. D.), ar525@cam.ac.uk (A. R.), robert.hoye@chem.ox.ac.uk (R. L. Z. H.)

**Abstract**

Polarised light is critical for a wide range of applications, but is usually generated by filtering unpolarised light, which leads to significant energy losses and requires additional optics. Herein, the direct emission of linearly-polarised light is achieved from light-emitting diodes (LEDs) made of $CsPbI_3$ perovskite nanoplatelet superlattices. Through use of solvents with different vapour pressures, the self-assembly of perovskite nanoplatelets is achieved to enable fine control over the orientation (either face-up or edge-up) and therefore the transition dipole moment. As a result of the highly-uniform alignment of the nanoplatelets, as well as their strong quantum and dielectric confinement, large exciton fine-structure splitting is achieved at the film level, leading to pure-red LEDs exhibiting a high degree of linear polarisation of 74.4% without any photonic structures. This work unveils the possibilities of perovskite nanoplatelets as a highly promising source of linearly-polarised electroluminescence, opening up the development of next-generation 3D displays and optical communications from this highly versatile, solution-processable system.

Linearly-polarised light is essential in numerous applications, including radiometric analysis, bio-imaging, counterfeit detection, liquid crystal displays, and emerging 3D display technologies[1-8]. Conventionally, linear polarisation is obtained by passing unpolarised light through a linear polariser, which is typically constructed from highly-oriented long-chain molecules or wire grids with photonic structures, reducing the intensity by a factor of two or more[1,8]. Organic molecules with aligned transition dipole moments (TDMs) can enable LEDs with linearly polarised electroluminescence (see Supplementary Table 1 for a comparison of reported materials), but the degree of polarisation (DOP) achieved until now have only been below 40% [9-11]. Photonic structures, such as Ag nanogratings, are generally needed to generate higher DOPs (see Supplementary Table 1)[12]. On the other hand, individual single particles of anisotropic inorganic nanostructures (e.g., cadmium selenide nanorods or indium phosphide nanowires) can achieve linearly polarised PL with DOP exceeding 70% (see Supplementary Table 1) [1,2,13-21]. However, the DOP values significantly decrease when integrating single particles together to form nanoparticle films without using patterned photonic substrates because the nanoparticles would then have a random orientation, which limits the ability to translate the polarised emission from single particles to the device level [1,18,22]. Therefore, approaches are needed to finely control the orientation of the anisotropic nanostructures in order to achieve high DOPs in films and devices.

Colloidal perovskite nanocrystals are a promising emerging class of light-emitting materials for display applications [23]. Importantly, it has been shown that perovskite nanocrystals can self-assemble [24,25]. Their bandgaps can be tuned over the entire visible to near-infrared wavelength ranges by controlling the composition and dimensionality[23,26,27]. Furthermore, perovskite nanocrystals benefit from defect tolerance in low-bandgap compositions, such as iodide-based perovskites[28], and can be passivated using a wide variety of approaches, enabling high photoluminescence quantum yields (PLQYs) reaching up to unity[24,25,29-31]. Halide perovskites can also be fabricated as anisotropic nanostructures, such as nanoplatelets and nanowires, which have been reported to yield high DOPs in PL[32-34], and are therefore promising for generating linearly polarised light. Becker *et al*. found that caesium lead halide nanocrystals have an exciton fine structure with distinct bright triplet states, and this could deliver strong linearly polarised emission[34,35]. Seminal works on controlling the transition dipole moment (TDM) alignment in weakly-confined $CsPbBr_3$ nanoplatelets (9±2 nm thickness, >10 monolayers) have recently appeared [36,37]. But as far as the authors are aware, there is no report on the orientation-controlled self-assembly of strongly-confined perovskite NPls (*i.e.*, NPls with thicknesses smaller than the exciton Bohr radius), and no demonstration of direct emission of high linearly-polarised EL from halide perovskite LEDs.

In this work, we overcome these pressing challenges to achieve linearly polarised EL directly from perovskite LEDs, which we demonstrate with colour-pure $CsPbI_3$ NPls. To do this, we focus on addressing four critical factors. The first is in devising a synthesis route to achieve $CsPbI_3$ NPls with a high degree of uniformity (necessary to achieve colour-pure emission), as well as strong quantum confinement (necessary to achieve exciton fine-structure splitting and polarised emission). The second is in self-assembling these $CsPbI_3$ NPls with aligned TDMs. We investigate the use of solvents with different evaporation rates on the preferred orientation of the NPls, and the uniformity in their preferred orientation. The third factor is in ensuring that the materials have exciton-dominated radiative recombination, and minimising the influence of surface defects on non-radiative recombination. This requires effective defect passivation to achieve high photoluminescence quantum yields (PLQYs) more efficient LEDs. The final factor is to achieve a sufficiently large exciton fine-structure splitting to give high EL DOPs. This requires not only strong quantum confinement, but also dielectric confinement in a cubic lattice.

To address these four factors, we utilised structural (transmission electron microscopy, TEM; grazing-incidence wide-angle X-ray scattering, GIWAXS), optical ($k$-space Fourier microscopy) and first-principles analyses to understand the self-assembly properties and dipole alignment of the films. We measured the PL spectra at 5.2 K to compare the fine-structure splitting energy of the strongly-confined NPls against a weakly-confined nanocube reference. We demonstrated pure-red $CsPbI_3$ LEDs with an EL DOP of 74.4%, which, to our knowledge, is the highest EL DOP for any organic or inorganic emitter reported thus far (Supplementary Table 1). The LEDs reported herein achieve an EQE of 2.7%, to our knowledge the highest for any strongly-confined perovskite emitter, in addition to being the first demonstration of pure-red EL from inorganic perovskite NPl LEDs.

**Uniform, strongly-confined $CsPbI_3$ nanoplatelets with colour-pure luminescence**

$CsPbI_3$ NPls were synthesised by the spontaneous crystallisation method using a nonpolar toluene solvent for $PbI_2$ precursors and oleic acid for Cs-acetate precursors (see Methods)[38].

The PL spectra and TEM images of the synthesised NPls are given in Fig. 1 and Supplementary Fig. 1. Previously-reported CsPbI$_3$ NPls had a shoulder in the PL[38]. To eliminate this and achieve colour-pure emission, we carefully investigated the effects of the precursors used, reaction temperature, and ratio in which the precursors were mixed, as shown in Supplementary Fig. 2a-c. We found that using caesium carbonate as a precursor led to a bimodal thickness distribution because carbonate reacts with the oleic acid ligands to form water that could damage NPls by removing surface ligands and cause agglomeration. This was addressed by switching to acetate as the spectator anion, as well as lowering the reaction temperature using an ice bath in order to reduce the growth rate of the NPls. In addition, by reducing the quantity of the caesium acetate precursor, we avoided NC formation (further details in Supplementary Note 1). The size distribution of the optimised NPls was analysed by TEM, confirming that the NPls have a length of 22±2 nm, width of 22±3 nm and a thickness of 2.6±0.5 nm (Supplementary Fig. 2d). The thickness of the NPls was much smaller than the Bohr radius of CsPbI$_3$ (~12 nm) [39], and therefore a strong excitonic peak was observed in the absorption spectra (Fig.1b-d). The centre of the PL was at approximately 600 nm (3 monolayer NPls; see spectra in Fig. 1 b-d), which is blue-shifted compared with weakly-confined CsPbI$_3$ NCs (emitting at 700 nm wavelength).

**Controlling the orientation of CsPbI$_3$ nanoplatelets through self-assembly**

We developed a strategy for NPl orientation control based on their self-assembly at the solid-air interface upon the controlled evaporation of solvents with different vapour pressures [40]. After centrifuging the crude solution at 22 769 RCF (*g*) (14 000 rpm) for 10 min, the precipitated

NPls were dispersed in solvents with different boiling points (bp) and vapour pressures ($p_v$): hexane (bp = 69 °C, $p_v$ = 20.49 kPa), cyclohexane (bp = 80.75 °C, $p_v$ = 13.01 kPa), and heptane (bp = 98.42 °C, $p_v$ = 6.08 kPa). The use of solvents with different bp and $p_v$ changes the evaporation rate of the solvents during spin-coating. As a result, the NPls can be kinetically trapped on the substrate to form films with different orientations without changing their dimensions, thus maintaining their absorption and PL profiles (Fig. 1b-d). For a slowly-evaporating solvent, such as heptane, the NPls adopted a face-down orientation (Fig. 1b). By contrast, using a fast-evaporating solvent, such as hexane, resulted in an edge-up orientation (Fig.1d). Additional TEM images with smaller magnification showing NPls with different orientations can be found in Supplementary Fig. 1.

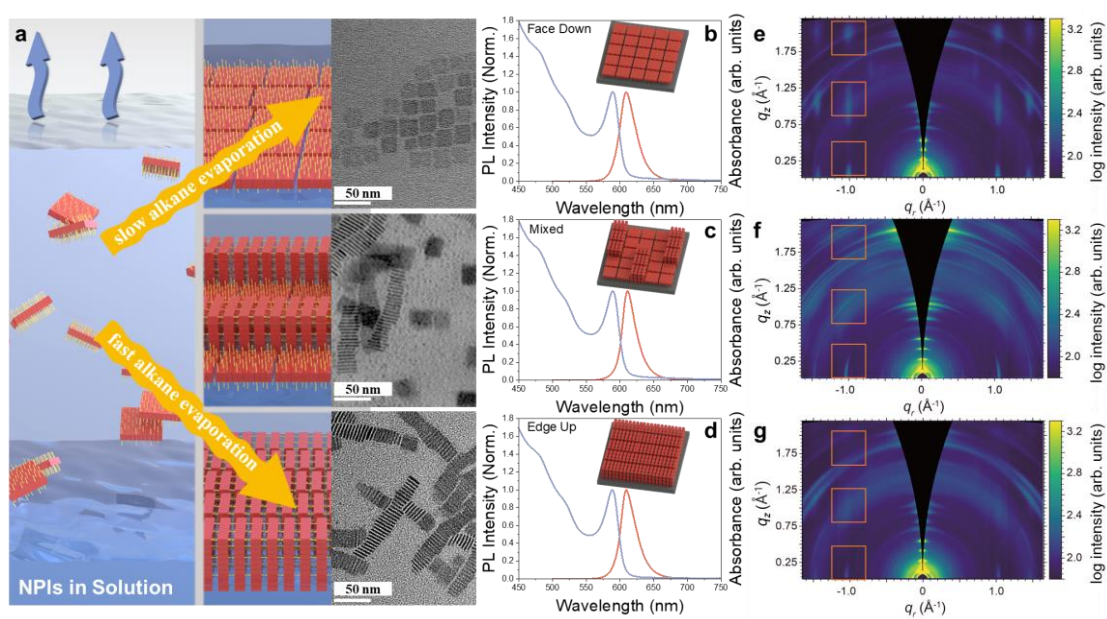

**Fig. 1 | Orientation control of self-assembled CsPbI$_3$ nanoplatelets. a,** Schematic illustrating how the orientation of the NPls could be controlled by adjusting the solvent evaporation rate. The insets show TEM images of the face-down (top), mixed (middle) and edge-up (bottom) NPls. **b-d,** Steady state PL and UV-Vis spectra of face-down, mixed, and edge-up NPl films, respectively. GIWAXS patterns of **e,** face-down, **f,** mixed and **g,** edge-up NPl films.

The change in NPl orientation was corroborated by GIWAXS measurements (Fig. 1 e-g). Films with face-down NPls (Fig. 1e) showed strongly observable peaks associated with (100), (101) and (102) planes (highlighted in the orange boxes from bottom to top, respectively), which became weaker and non-existent in the mixed (Fig. 1f) and edge-up (Fig. 1g) films, respectively [41]. In addition, the spread in the scattering of X-rays perpendicular to the diffraction rings in Fig 1e confirms the layer-by-layer stacking of NPls in the face-down orientation. Vertical stacking of the NPls when they were face-down resulted in more repetitive planes than if they were edge-up, leading to stronger diffraction from these periodic planes; please refer to the intensity difference of the fringe inside the orange box for Fig. 1e and Fig. 1g [41].

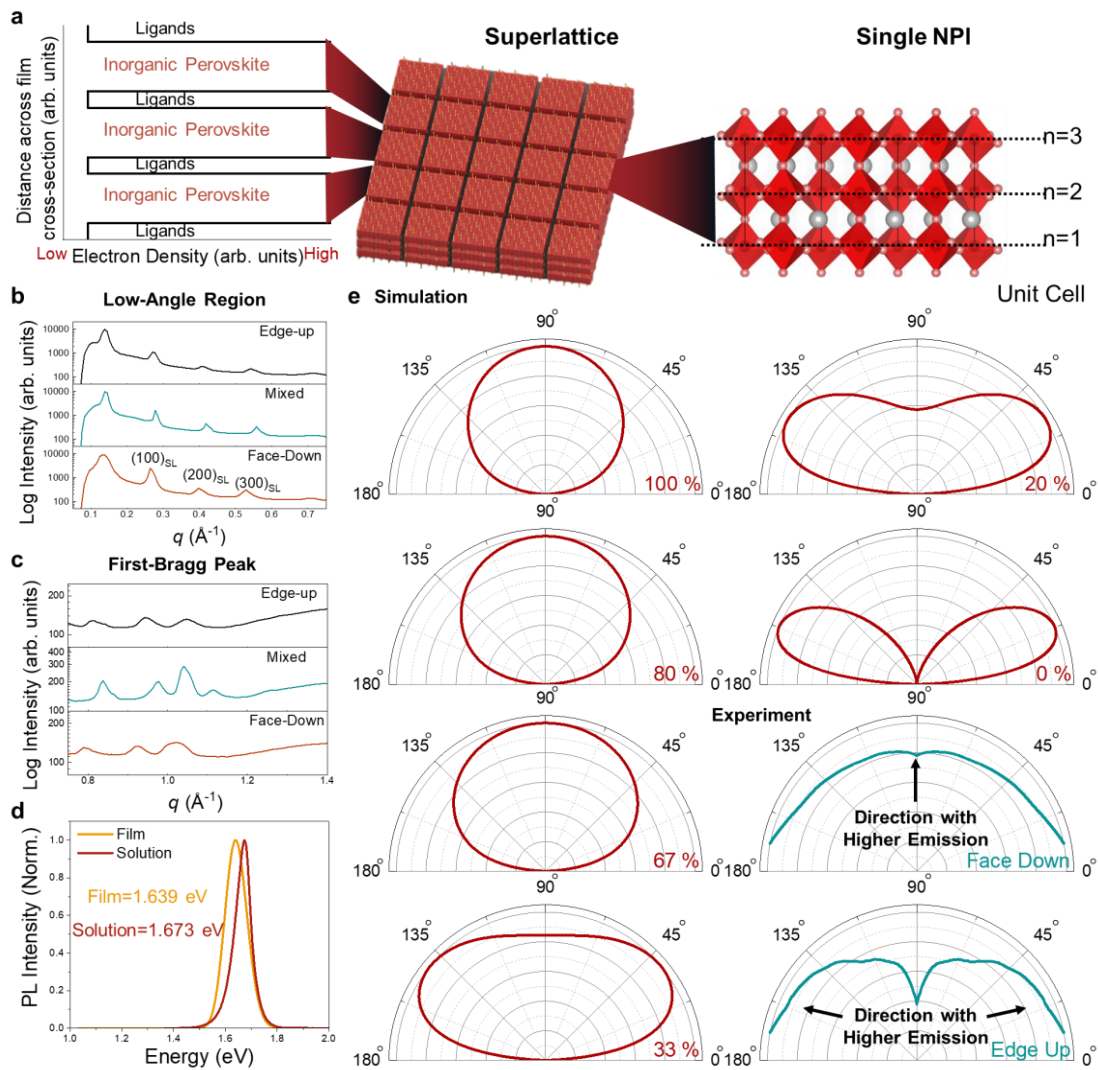

**Fig. 2 | Structural and optical characterisation of self-assembled NPl films. a,** Illustration of superlattices formed from the stacking of multiple NPls, and how the electron density changes between inorganic perovskites and organic ligands. **b,** Line cut of the low angle region

in the GIWAXS measurements which demonstrates the formation of a superlattice. **c,** Line cut of the first Bragg peak region in the GIWAXS measurements. **d,** The PL of the NPls in solution and films. The red shift in PL from solution to film could be caused by the self-assembled structures, as the NPls are stacking together, the thickness or electronic coupling might be altered. **e,** Simulated (red) and experimental (blue) outcoupling from NPl films. Simulated light outcoupling of the NPls films with different percentages of horizontal transition dipole moment (100% = all NPls face-down, whereas 0% = all NPls edge-up) to represent different degrees of preferred orientation of the NPls, assuming the outcoupling contribution of the TDM in the confined direction is negligible. The simulated radiation patterns are integrated over the azimuth coordinate as a function of the polar angle (see Methods for details on the simulations). Experimental measurements of the angle-dependent PL obtained from *k*-space Fourier Microscopy are shown in blue. Details on data collection and analysis can be found in Supplementary Fig. 3. 0° is where the collection is parallel to the substrate, and 90° is where the collection is normal to the substrate.

The analysis of line cuts in the GIWAXS data reveals that the high degree of NPl orientation is related to the formation of superlattices during film formation. A schematic of a NPl superlattice is shown in Fig. 2a, depicting face-down NPls stacked vertically and separated by organic ligands. The regular variation in electron density across the cross-section of the superlattice between the inorganic sub-lattice (high electron density) and organic ligands (low electron density), as illustrated in Fig. 2a, results in peaks in the low-angle scattering region from the GIWAXS scans (Fig. 2b) [36,41]. Another observation consistent with the formation of superlattices is the PL of the films having lower energy than those of $CsPbI_3$ NPls in colloidal solution, as shown in Fig. 2d. This can be explained by two possible mechanisms. First, strong alignment of transition dipole moments in the strongly-confined NPls (as occurs in superlattices) leads to lower-energy delocalised emissive states. Second, an increase in the dielectric constant of the surroundings around each NPl in superlattices (due to the presence of many high-dielectric constant NPls) compared with isolated NPls in solution, reduces the dielectric confinement[42,43]. The red-shift in the PL of NPls in films *vs.* colloidal solution was approximately 34 meV, which matches with previous reports of PL red-shifts in superlattices of $CsPbBr_3$ and $CsPbBr_{0.72}I_{0.28}$ nanocrystals (12-96 meV) compared with their isolated NCs.[42,43]

Apart from structural evidence for self-assembly, optical analysis also provides strong evidence. Optical simulations of light emission from the films with different transition dipole alignments are shown in Fig. 2e. The labels inset in these red contour plots (100% to 0%) represent the percentage of TDMs that are horizontally aligned. These simulations showed that when the dipole was aligned out-of-plane (*i.e.*, 0% horizontal TDM, or perfectly edge-up NPls), the vertical emission profile mainly came from the out-of-plane dipole (*i.e.*, along the thickness direction), leading to a lower PL intensity in the normal position compared to the horizontal direction. In contrast, when the NPls were perfectly face-down (*i.e.*, 100% horizontal TDMs), the emission normal to the plane mainly came from the horizontal dipole, leading to stronger PL emission in the normal direction. Changing the NPl orientation from edge-up to face-down would therefore lead to an increase in the normal PL intensity due to an increase in the percentage of horizontal transition dipole moments contributing to this emission.[15,36,37,44]

To validate the simulation results, angle-resolved PL was extracted from *k*-space Fourier microscopy measurements (Fig. 2e – blue plots; experimental details in Supplementary Fig. 3 and Supplementary Note 2). In the experimental results, 90° is where the collection was perpendicular to the substrates, and 0° is collection parallel to the substrates. The PL spectra at 90° and 11° can be found in Supplementary Fig. 3b and c. By integrating the PL spectra at each collection angle, we found that the edge-up NPl film had higher PL intensities at low collection angles, which then decreased markedly as the collection angle increased. By contrast, the PL intensity for the face-down NPl films remained relatively constant over the range of collection angles (Fig. 2e and Supplementary Fig. 3c). These trends in angle-resolved PL agree with the simulation results. By sampling across five points in each film to eliminate random error, the same trends shown in Fig. 2e were observed. We note that the measured angle-resolved PL for the face-down NPls had higher intensity at collection angles close to the substrate than expected from the simulations for 0% horizontal TDMs. This was because we assumed in the simulations that the TDMs in the confined direction (out-of-plane dipoles) to have negligible contribution to the emission profile. But when face-down NPls stacked vertically to form superlattices, the emission from the out-of-plane dipoles would accumulate and lead to a non-negligible contribution towards the emission profile, thus accounting for more isotropic PL intensities observed over all collection angles. Moreover, it is difficult to accurately determine the percentage of in-plane TDMs using the current data, as the collection solid angle covered a few steradians, and the data was therefore averaged over several NPls. Nevertheless, these data collectively suggest that the films have higher vertical outcoupling from face-down NPls than edge-up NPls.

The differences in light outcoupling with NPl orientation indicate that there was a substantial degree of dipole alignment in the films, which is beneficial for generating polarised emission[15]. For example, Achtstein and co-workers illustrated that single-particle NPls have stronger directional emission with high linear polarisation than isotropic spherical quantum dots [44]. Also, Gao and co-workers showed that aligning TDMs was critical to achieve polarised PL from MAPbBr$_3$ polycrystalline films[45]. Therefore, the demonstration of fine control over TDM alignment through the formation of self-assembled superlattices is highly promising for achieving polarised light emission.

**Reducing non-radiative recombination through surface passivation**

Before developing these self-assembled NPl films into LEDs, we first identified and reduced loss processes at NPl interfaces. We performed X-ray photoemission spectroscopy (XPS) measurements on the perovskite nanoplatelets, and found the surface to be terminated by Pb and I (Supplementary Fig. 4), with noticeable quantities of metallic Pb (Pb$^0$). These Pb$^0$ species, as well as possibly iodide vacancies, could limit the PLQY to 48±2% (Fig. 3a). We found that we could improve the PLQY to 65±1% by passivating the NPls with PbI$_2$ coordinated with oleate and oleylamine ligands. The PbI$_2$ passivators could coordinate with Pb$^0$ and fill I vacancies to reduce the density of recombination centres, especially at the surface. Full details are given in Supplementary Note 3.

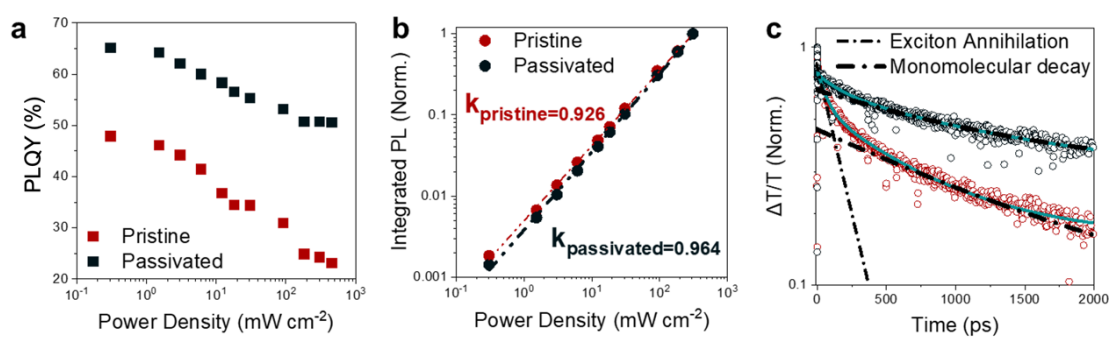

**Fig. 3 | Photoluminescence quantum yield (PLQY) of pristine and passivated NPls. a,** Power-density dependent PLQY (405 nm CW laser excitation). **b,** Power-density ($P$) dependence of the integrated PL peak intensity. $I_{PL} \propto P^k$, where $k$ is the exponent. When $k$ is close to 1, radiative recombination proceeds in a monomolecular manner, consistent with the behaviour of excitons. **c,** Short timescale (0–2 ns time window) transient absorption spectroscopy measurements of the ground state bleach. Excitation was with a 400 nm wavelength pump laser, 301.5 µJ cm$^{-2}$ pulse$^{-1}$ fluence, and 595 nm wavelength probe. All samples were in colloidal solution, measured in a cuvette with a 1 mm thick cavity.

As shown in Fig. 3a, the PLQY of the NPls decreased with an increase in the excitation power density regardless of whether they were passivated. This trend contrasts to weakly-confined nanocrystal systems, where the PLQY usually increases with excitation power due to trap filling[46]. The decrease in the PLQY of the NPls with increasing power density was due to bimolecular exciton-exciton annihilation. Fig. 3b demonstrated a linear relationship between the PL intensity and excitation power density, indicating that we have a exciton dominated recombination system, which arises from the high exciton binding energies (243 meV; see later in Fig. 5f) due to strong dielectric and quantum confinement[47,48].

To understand in greater depth the effect of passivation on exciton kinetics, we measured time-resolved PL (TRPL) and transient absorption spectroscopy (TAS) of the NPl solutions. The governing equations and our method for determining the rate constants are discussed in Supplementary Note 3, with details in Supplementary Fig. 4–6. The fitting results are shown in Supplementary Table 2. This analysis shows that after $PbI_2$-ligand passivation, the monomolecular exciton radiative recombination rate ($k_{rad}$) decreased from 0.008 $ns^{-1}$ (pristine) to 0.004 $ns^{-1}$ (passivated). Meanwhile, the non-radiative monomolecular exciton trapping rate ($k_{trap}$) decreased from 0.009 $ns^{-1}$ (pristine) to 0.002 $ns^{-1}$ (passivated). Taken together, these results show that the recombination process in the NPls became increasingly dominated by radiative processes after passivation (consistent with the PLQY measurements), and that the radiative lifetime of the excitons became longer. In addition, the non-radiative bimolecular exciton-exciton annihilation (EEA) rate decreased from 0.382 $cm^2\ s^{-1}$ (pristine) to 0.282 $cm^2\ s^{-1}$ (passivated). This can be seen from Fig. 3c for passivated and pristine NPl films. The dotted line indicates the mono-exponential decay of the passivated films. When super-positioning the fitted bimolecular EEA and monomolecular decay of the passivated sample to the pristine sample in Fig. 3c, there is a clear mismatch of the fitting to the experimental data of the pristine sample, confirming that there were a trap-induced carrier losses at early timescales after excitation for the samples without passivation.

**Linearly polarised LEDs with orientation-controlled $CsPbI_3$ nanoplatelets fulfilling Rec. 2020**

Edge-up and face-down NPl films were fabricated into LEDs using the following structure: ITO//PEDOT:PSS/poly-TPD/NPls/BCP/Ag (Fig. 4b inset). The EL spectra at different voltages (3.5-7 V) are shown in Fig. 4a. A red emission with CIE coordinates of (0.661, 0.338) was achieved, which is close to the Rec. 2020 standard for pure-red emission (Supplementary Fig. 7a). Unlike the mixed Br-I perovskite systems, the composition-purity of $CsPbI_3$ NPls in the halide site enabled the devices to avoid halide segregation even at high applied biases (7 V) [46,49]. It was only after increasing the applied bias from 3 V to well above the operational voltage with peak EQE that any changes in the EL spectrum became noticeable. At 7 V, the CIE coordinates marginally changed from (0.661, 0.338) at 3 V to (0.658, 0.341). This was due to the thermally-induced merging of NPls due Joule heating from the high current density (approximately 1 mA $cm^{-2}$) by this point. Through NPl orientation control, we increased the average EQE of the pristine devices (control) from 1.12% to 1.22% when changing the NPl orientation from edge-up to face-down by improving outcoupling (refer back to Fig. 2e). After adding the effects of surface passivation using $PbI_2$-ligand, the average EQE increased from 1.12% to 1.67% (edge-up NPls), and from 1.22% to 2.38% (face-down NPls) (Supplementary

Fig. 7b). The current injected was reduced after passivation, especially for the face-down NPl devices (Supplementary Fig. 7c). The champion NPl device (face-down NPls with PbI$_2$-ligand passivation) achieved 2.7% EQE at 4 V (1.25 mA cm$^{-2}$), with a maximum luminance of 408 cd m$^{-2}$ obtained at 6V (49.95 mA cm$^{-2}$) (Fig. 4b, Supplementary Fig. 7d).

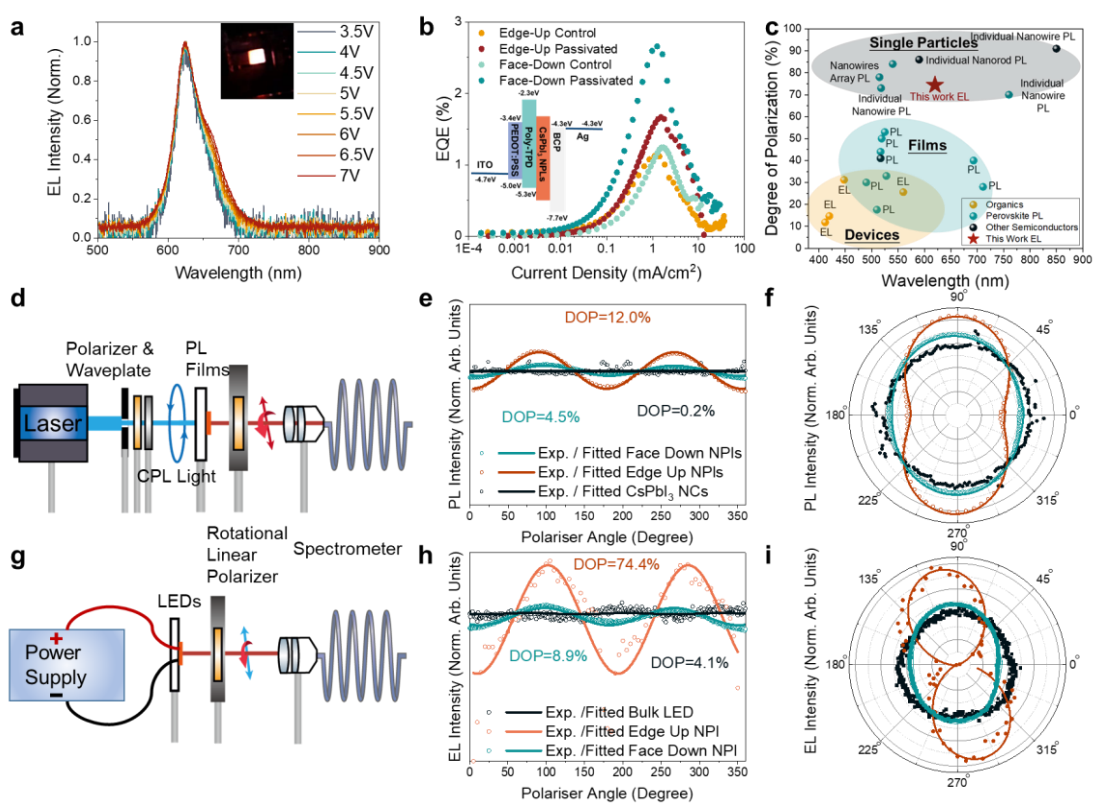

**Fig. 4 | Linearly polarised emission from CsPbI$_3$ NPl films and devices. a,** Electroluminescence (EL) spectra of optimised NPl LEDs from 3.5 V to 7 V. Inset shows the LED under operation (pixel area 0.16 cm$^2$). **b,** Current density against EQE for pristine and passivated edge-up and face-down NPl LEDs. Inset, the optimised LED device structure: ITO/PEDOT:PSS/poly-TPD/NPls/BCP/Ag. **c,** Summary of degree of polarisation of different materials at single particle, film and device level. **d,** Schematic of producing and testing the polarised photoluminescence. **e,** Polarisation dependence of the PL for face-down, edge-up and bulk FAPbI$_3$ films at a fixed detection angle of 0° with respect to the substrate normal. **f,** Polar plots of the polarised PL. **g,** Schematic of producing and testing the polarised electroluminescence. **h,** Polarisation dependence of the EL for edge-up and bulk FAPbI$_3$ LEDs at a fixed detection angle of 0° with respect to the substrate normal. **i,** Polar plots of the polarised EL. The degree of polarisation, DOP, was calculated from: $\text{DOP} = (I_{max} - I_{min})/(I_{max} + I_{min})$

The self-assembly of strongly-confined, and highly orientated NPls led to strongly linearly polarised EL from the LEDs, as shown in Fig. 4c-i. To measure the DOP, circularly polarised light (CPL) was generated by passing unpolarised light from the laser LEDs into a linear polariser and a waveplate. The emission of the sample was then passed through a linear polariser mounted onto an electrically driven rotation stage and coupled into a spectrometer. (Fig. 4d). The degree of polarisation of the EL was recorded in the same way (Fig. 4g). Both PL and EL showed clear polarised emission relative to more isotropic emitters, such as weakly-confined $CsPbI_3$ nanocube films and bulk $FAPbI_3$ thin film LEDs.

For the measurements of the DOP in PL (Fig. 4e), we found that the edge-up configuration (DOP=12%, PL) had a stronger degree of polarisation than the face-down NPls (DOP=4.5%, PL), which is consistent with previous reports of self-assembled CdSe NPls[40], as well as with theoretical predictions[35] (Fig. 4f). By contrast, the isotropic $CsPbI_3$ NC control sample had a low DOP of only 0.4% PL. To measure the DOP for EL, we accounted for changes in EL intensity over time under operation (Supplementary Fig. 7e and f) in order to extract only the changes in EL with the angle of the rotational linear polariser (Fig. 4g) over several cycles. For the edge-up LEDs, an EL DOP of 74.4% was achieved (Fig. 4h and i), which is the highest reported for any organic or inorganic LED without using photonic gratings (Supplementary Table 1 and Fig. 4c) [18,50,51]. Notably, the DOP achieved in these film-based LEDs reached the single-particle level (Fig. 4c). In the next section, we discuss why the DOP for EL could exceed that for PL.

**Strong degree of linear polarisation originating from large exciton fine-structure splitting**

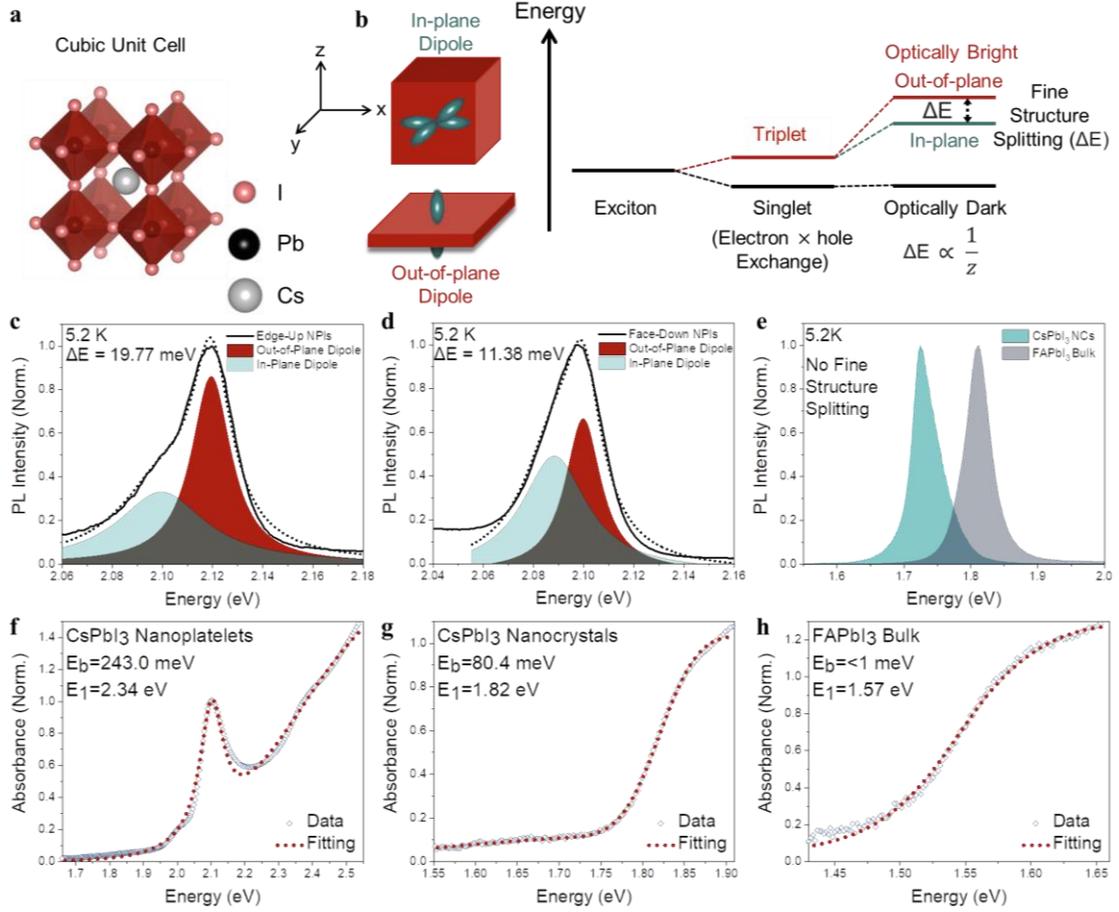

**Fig. 5 | Origin of linearly polarised emission. a,** Crystal structure of cubic (α-phase) CsPbI$_3$ ($Pm\bar{3}m$). **b,** Schematic diagram to illustrate the influence of quantum and dielectric confinement on exciton fine structure splitting (Δ$E$). The z-axis is across the thickness of the NPls. Low temperature (5.2 K) PL spectra of **c,** edge-up and **d,** face-down NPl films. The fitted peaks represent the emission from the exciton fine structure. In these measurements, the detector was held perpendicular to the film. **e,** Low temperature (5.2 K) PL spectra of CsPbI$_3$ nanocubes and bulk FAPbI$_3$ thin films, from which no excitonic fine-structure splitting was observed. **f-g,** Elliot model fitting of the absorbance of CsPbI$_3$ NPls, NCs and bulk FAPbI$_3$ to extract the exciton binding energy ($E_b$). $E_1$ is the first excited state. The fitting model used is based on Ref. 52 (fitting code from Ref. 53).

To understand the origin of the high DOP in the NPl films and devices, we measured the PL of the face-down and edge-up CsPbI$_3$ NPl films (Fig. 5a), and compared with weakly-confined CsPbI$_3$ NC films and FAPbI$_3$ bulk thin films at 5.2 K. Linearly polarised PL can arise from the fine-structure splitting (FSS) of band-edge excitons [2,35]. In lead-halide perovskites, short- and long-range electron–hole (e-h) exchange interactions result in a splitting of the band-edge excitonic states into an optically inactive singlet state with total angular momentum $j = 0$, and three optically active triplet states with $j = 1$ (Fig. 5b)[54]. The triplet states can be split further into two or more energetic states in tetragonal or orthorhombic perovskites through crystal field effects[54], or in cubic perovskites through strong quantum and dielectric confinement[35,55-57]. For example, Ghribi *et al.* showed through computational analyses that whilst non-confined cubic perovskites have degenerate optically-active triplets, the state with an out-of-plane dipole splits away from the two in-plane dipole states as the thickness of these nanocubes are reduced, and strongly quantum and dielectric confined nanoplatelets are realised[34] (see Fig. 5b for an illustration of this effect). Indeed, groups have reported the experimental observation of FSS from single particle $CsPbBr_3$ NPls, *i.e.*, the PL spectra split into multiple peaks (1-3) at cryogenic temperature[20,58]. However, the FSS reported was weak, and became especially difficult to observe at the film level when the NPls or NCs were randomly distributed. In contrast to these previous experimental results, we were able to clearly observe PL splitting at 5.2 K at the film level (Fig. 5c and d), which may have been due to the uniform orientation of the NPls and the alignment of their transition dipole moments through self-assembly. We extracted an FSS energy of 19.77 meV (edge-up) and 11.38 meV (face-down) by fitting the PL spectra we measured at 5.2 K. These values are significantly larger than the FSS energy of the reported isolated cuboidal NC (~1 meV)[20,34]. Since our NPls are strongly confined in the *z*-direction, the aspect ratio (thickness to length/width ratio) is very small (~0.13), and the FSS values we obtained for such small aspect ratio NPls match well with predictions from previous theoretical analyses (~20 meV) [35].

Compared with the strongly-oriented NPl films we achieved here, no obvious FSS-induced PL splitting or polarised light emission were observed in weakly-confined NCs or bulk thin films (Fig. 5e). There are two possible reasons for this. Firstly, there is the lower exciton binding energy ($E_b$) in the NCs (80.4 meV) and bulk films (negligible) than in the NPls (243 meV), as shown in Fig. 5f-h. All samples have the same cubic phase, so structural effects would have minimal impact compared to the effects of $E_b$. The smaller $E_b$ in the NCs and bulk films would result in 1) radiative recombination not being solely due to excitons, and 2) a significantly reduced or negligible FSS. Secondly, there were no aligned TDMs in the NCs and bulk films, in contrast to the NPls. Supplementary Fig. 9 shows the XRD patterns of drop-cast NPls and NCs on glass substrates. Despite both of them having the same cubic phase, strong satellite peaks were only observed for the NPl films, which is consistent with superlattice formation[59]. The lack of superstructures for NCs implies a more random orientation of TDMs in the films, thereby leading to no well-defined polarised emission.

We were able to fit two peaks to the low-temperature PL spectra of the NPls (Fig. 5c,d). We attributed the higher energy PL peak to emission from the out-of-plane dipole (*z*-direction), and the lower-energy PL peak from the in-plane dipoles (*x* and *y* directions). The emission from each fine-structure state exhibits single polarisation, with the *x*- and *y*-direction polarisation being orthogonal to each other. Since we have similar sizes in the x- and y directions, these two states are essentially degenerate in α-CsPbI$_3$, and the overall emission would have negligible polarisation as they cancel each other out. Therefore, the linearly polarised emission from the NPls would come from the higher energy state (*z*-direction dipole). From Fig. 5c,d, it can be seen that the edge-up films have stronger emission from these higher-energy states. This was because the collection angle is normal to the film, and the emission in this direction has a strong contribution from the aligned out-of-plane dipole, which has higher energy level in the fine structure. This could explain why the edge-up film have a higher DOP than the face-down NPl films. However, the total DOP and PL splitting should be influenced by a combination of depolarisation within the FSS manifold due to energy transfer[35,60], the detection geometry relative to the crystal plane[58] and the local field effect[61].

The EL DOP (74.4%) significantly exceeded the PL DOP (12.0%), and this could be explained by further increases in dipole alignment by the applied electric field during LED operation that leads to most of the injected carriers recombining at one particular energy level with minimal energy transfer to other triplet states. This observation matches well with previous studies on GaAs quantum dots, in which an increase in the DOP was also observed after increasing the applied electrical field [62].

**Conclusion**

In conclusion, we have made the first demonstration of the direct generation of linearly-polarised electroluminescence from halide perovskites through use of strongly-confined $CsPbI_3$ nanoplatelets. These nanoplatelet LEDs also fulfil Rec. 2020 and have 2.7% EQE, the highest reported thus far for strongly-confined NPl LEDs (Supplementary Fig. 8)[21]. The orientation of the NPls, and hence the percentage of in-plane/out-of-plane transition dipole moments, can be manipulated by the evaporation rate of the solvent the colloidal NPls are dispersed in. Both structural (TEM, GIWAXS) and optical (angular-resolved PL) characterisation confirmed our ability to control the orientation of NPl films through self-assembly. Detailed carrier dynamics measurements using time-resolved photoluminescence and transient absorption spectroscopy helped us to develop a passivation process for CsPbI$_3$ NPls that reduced both monomolecular trap-assisted-recombination and bimolecular exciton-exciton annihilation. Importantly, we demonstrated a high degree of linear polarisation (74.4%) in the EL from pure-red LED devices, which is attributed to three key factors: 1) highly uniform alignment of the NPls, 2) high PLQYs, such that the emission is strongly governed by the radiative recombination of band-edge excitons, and 3) large exciton fine structure splitting due to strong quantum and dielectric confinement, which was preserved at the film-level due to the formation of strongly-aligned NPl superlattices. This work opens up a new frontier in enabling applications for displays and optical communications with more efficient linearly polarised LEDs.

**Acknowledgements**

J.Y. and R.L.Z.H. acknowledge support from a UK Research and Innovation (UKRI) Frontier Grant (no. EP/X029900/1), awarded via the European Research Council Starting Grant 2021 scheme. J.Y. also gives thanks to Cambridge Philosophical Society for the Research Studentship Grant and Churchill College for various travel and research grant. R.L.Z.H. thanks the Royal Academy of Engineering through the Research Fellowships scheme (no. RF\201718\1701), as well as the Centre of Advanced Materials for Integrated Energy Systems (CAM-IES; EPSRC grant no. EP/T012218/1). T.K.B. gives thanks to the Centre for Doctoral Training in New and Sustainable Photovoltaics (EP/L01551X/2) and the NanoDTC (EP/L015978/1) for financial support. L.D. thanks the Cambridge Trusts and the China Scholarship Council for funding. M.A.R. and P.M.-B. acknowledge funding from the Deutsche Forschungsgemeinschaft (DFG, German Research Foundation) within Germany′s Excellence Strategy, EXC 2089/1–390,776,260 (e-conversion) and by TUM.solar in the context of the Bavarian Collaborative

Research Project Solar Technologies Go Hybrid (SolTech). R.A. acknowledges support from the Rutherford Foundation of the Royal Society Te Apārangi of New Zealand, the Winton Programme for the Physics of Sustainability, and Trinity College Cambridge. J.J.B. acknowledges support of ERC grant PICOFORCE (883703). K. Z. would like to acknowledge the EPSRC Centre for Doctoral Training in Graphene Technology (EP/L016087/1) for studentship. K.Z. and L.T.M would like to acknowledge EP/ P030467/1 equipment grant for the generation of microscopic data. The authors would like to thank Dr. Mark Isaacs for the collection of the X-ray photoelectron (XPS) data at the EPSRC National Facility for XPS ("HarwellXPS"), operated by Cardiff University and UCL, under Contract No. PR16195. This publication is part of the project NanoLEDs with project number 17100 of the research programme High Tech Systems and Materials which is (partly) financed by the Dutch Research Council (NWO). S.D.S. acknowledges the Royal Society and Tata Group (grant no. UF150033). The work has received funding from the European Research Council under the European Union's Horizon 2020 research and innovation programme (HYPERION, grant agreement no. 756962; PEROVSCI, 957513). The authors acknowledge the EPSRC (EP/R023980/1, EP/S030638/1, EP/V061747/1) for funding. L.P. acknowledges support from the Spanish Ministerio de Ciencia e Innovación through Ramón y Cajal grant (RYC2018-026103-I) and the Spanish State Research Agency (Grant PID2020-117371RA-I00) and a grant from the Xunta de Galicia (ED431F2021/05).

**Author contributions**

J.Y. and R.L.Z.H. conceived of the project. L.D. contributed to part of the project design under the supervision of N.C.G and S.D.S. J.Y. J.S. and L.P. synthesised the materials and films. J.Y. and Y.H. performed the transient absorption spectroscopy measurement. J.Y., A.R., J.W. and W.Z. optimised devices. T.B. performed the polarisation measurements. R.G., J.E.H, M.S. S.V.R, and P.M.B. performed the GIWAXS measurement and analysis. D.P., G.G. and A.F. K. constructed the optical simulation model. J.Y. and S.G. performed the low temperature PL measurement. J.H., R. A and J.J B performed and analysed the *k*-space Fourier microscopy measurement. Y.S. supplied the $FAPbI_3$ LEDs. K.Z. and L.T.M performed the TEM measurements. A.R. and R.L.Z.H. supervised the work. All authors contributed to writing and editing the manuscript.

**Conflicts of Interest**

Authors declare no competing interests.